# MULTIPLAYER QUANTUM GAMES AND ITS APPLICATION AS ACCESS CONTROLLER IN ARCHITECTURE OF QUANTUM COMPUTERS

PAULO BENÍCIO DE SOUSA and RUBENS VIANA RAMOS

*Departamento de Engenharia de Teleinformática – Universidade Federal do Ceará - DETI/UFC*
*C.P. 6007 – Campus do Pici - 60755-640 Fortaleza-Ce Brasil*

One of the basics tasks in computer systems is the control of access of resources. Basically, there is a finite amount of resources that can be, for example, the CPU, memory or I/O ports, and several processes requiring those resources. If there is not enough resource in order to attend the demand, some kind of control access has to be employed. In this work, recognizing the resource sharing problem as a competition, we employ a simplified multiplayer quantum game as an access controller. The proposed quantum game can be employed in the architecture of quantum computers.

## 1. Introduction

Independently of the technology used to be built, quantum computers, as happens with classical computers, will need an architecture able to manage resource like I/O ports, memory, CPU and their data exchange. Since those resources are finite and they are requested by different processes simultaneously, it is necessary to control the access to them in a fair way. In classical computers, the resources management is performed by structures named request queues where the processes requirements are stored and executed by the operational system in a FIFO (First In First Out) way. Regarding quantum computers, recognizing that the scenario where several processes are trying to use, for example, the CPU, simultaneously, is a kind of competition, we apply quantum multiplayer games in order to control the resource sharing. Using quantum games, the process that will have access to the resource in a specific moment, is the process that wins the turn of the game.

This paper is outlined as follows: in Section 2, a simplified quantum game model is proposed; in Section 3, the multiplayer game for access control is introduced; in Section 4, it is shown a simple quantum computer architecture using the proposed quantum game; at last, the conclusions are presented in Section 5.

## 2. Quantum Games – Simplified Version

Quantum games were introduced by Meyer [1] and Eisert [2] as a generalization of classical games [3].

The most famous quantum game is the quantum prisoner's dilemma, QPD. The Meyer-Eisert's model for the realization of this quantum game is as shown in Fig. 1.

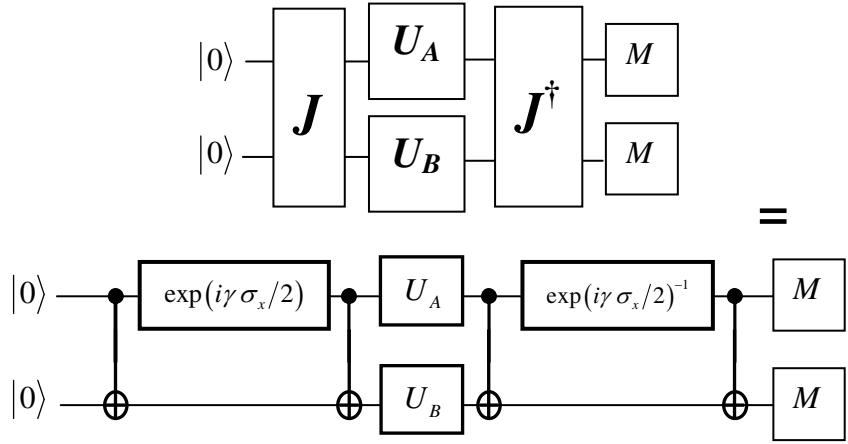

Fig. 1 – Quantum circuit for the quantum prisoners' dilemma. The single qubit gates $U_A$ and $U_B$ are the strategies and $M$ is a measurer.

The entangler matrix $J$ is of the type $J=\exp(i\gamma\sigma_x\otimes\sigma_x)$ and it operates as $J|00\rangle=\cos(\gamma/2)|00\rangle+i\sin(\gamma/2)|11\rangle$, hence, $0<\gamma<\pi$. The maximal entanglement is achieved when $\gamma=\pi/2$. The strategies are the choices for the single-qubit gates $U_A$ and $U_B$, whose general form is

$$U(\theta,\phi,\varphi)=\begin{bmatrix} e^{-i\phi}\cos(\theta/2) & -e^{i\varphi}\sin(\theta/2) \\ e^{-i\varphi}\sin(\theta/2) & e^{i\phi}\cos(\theta/2) \end{bmatrix} \quad (1)$$

In despite of the correctness and usefulness of Meyer-Eisert's model, other quantum game models are also possible [4], as the simplified model that we propose here and it is shown in Fig. 2.

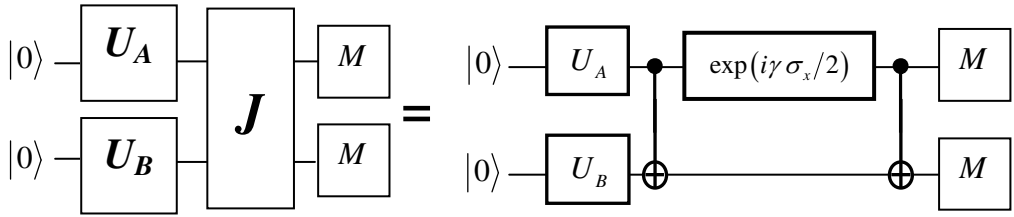

Fig. 2 – Simplified quantum game model.

In order to show that the quantum circuit presented in Fig. 2 realizes, in fact, a quantum game, one must prove that the most important property of a quantum game, the interference between strategies, is present. For this, let us consider a simple particular case where $\gamma=\pi/2$ and the players A and B choose their strategies using, respectively, the following single-qubit gates

$$U_{A(B)} = \frac{1}{\sqrt{2}} \begin{bmatrix} e^{i\phi_{A(B)}} & -e^{i\psi_{A(B)}} \\ e^{-i\psi_{A(B)}} & e^{-i\phi_{A(B)}} \end{bmatrix}. \tag{2}$$

In this case, the probabilities of the results 00 ($p_{cc}$), 01 ($p_{cd}$), 10 ($p_{dc}$) e 11 ($p_{dd}$) being obtained in the measurements, are given by

$$p_{cc} = 1/4\left[1+\sin\left(\left(\phi_B+\phi_A\right)+\left(\psi_B+\psi_A\right)\right)\right] \tag{3}$$

$$p_{cd} = 1/4\left[1-\sin\left(\left(\phi_B-\phi_A\right)+\left(\psi_B-\psi_A\right)\right)\right] \tag{4}$$

$$p_{dc} = 1/4\left[1+\sin\left(\left(\phi_B-\phi_A\right)+\left(\psi_B-\psi_A\right)\right)\right] \tag{5}$$

$$p_{dd} = 1/4\left[1-\sin\left(\left(\phi_B+\phi_A\right)+\left(\psi_B+\psi_A\right)\right)\right] \tag{6}$$

where $p_{cc}$ is the probability of A and B cooperate, $p_{cd}$ is the probability of A to cooperate and B to defect, $p_{dc}$ is the probability of A to defect and B to cooperate and, at last, $p_{dd}$ is the probability of A and B defect. As can be observed in (3)-(6), $0 \leq p_{cc}, p_{cd}, p_{dc}, p_{dd} \leq 1/2$, $p_{dd}=1/2-p_{cc}$ and $p_{cd}=1/2-p_{dc}$. Thus, depending on the choices of A and B, one of the results can never happen, showing a destructive interference between strategies. For the prisoners' dilemma, for example, if $\phi_B+\phi_A+\psi_B+\psi_A=\pi/2$, then the Nash equilibrium (both deffect) will never happen. In order to show more clearly the difference between the models presented in Figs. 1 and 2, we plot in Figs. 3-5 the possible average values of payoffs for the quantum prisoner's dilemma having the playoff table shown in Table 1. In the simulations shown in Figs. 3-5, $\gamma=\pi/2$ and the angles not mentioned are equal to zero.

| A \ B | $|C\rangle$ | $|D\rangle$ |
|---|---|---|
| $|C\rangle$ | (3,3) | (0,5) |
| $|D\rangle$ | (5,0) | (1,1) |

Table 1 – Payoff table for quantum prisoners' dilemma.

The average value of payoff for players *A* and *B* is given by:

$$\langle \$_k \rangle = \$_{CC,k} |\langle \Psi | CC \rangle|^2 + \$_{CD,k} |\langle \Psi | CD \rangle|^2 + \$_{DC,k} |\langle \Psi | DC \rangle|^2 + \$_{DD,k} |\langle \Psi | DD \rangle|^2, \quad (7)$$

where $k \in \{A,B\}$ and $|\Psi\rangle$ is the quantum state just before the measurements. The values $\$_{XY,Z}$ correspond to the payoffs shown in Table 1. For example, $\$_{CD,A}=0$ and $\$_{CD,B}=5$.

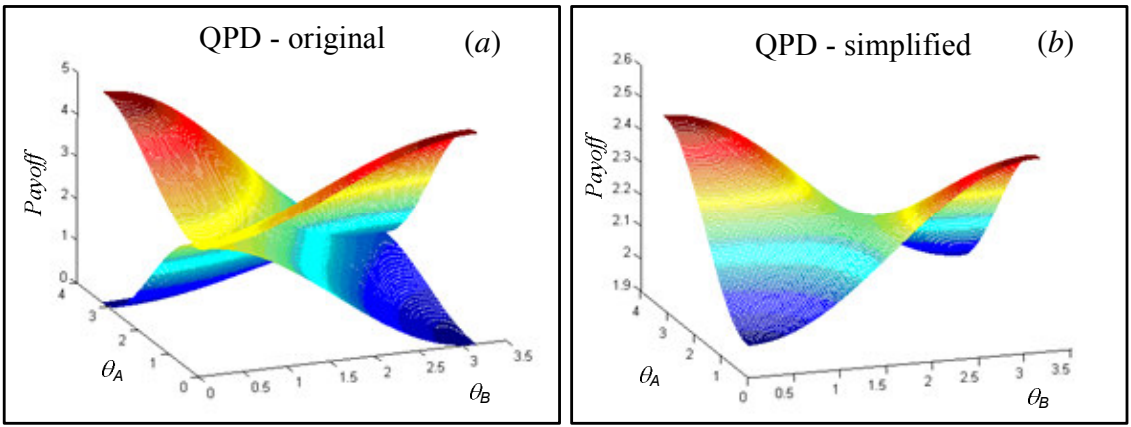

Fig. 3 – Average values of payoffs for players *A* and *B* versus $\theta_A$ and $\theta_B$, using Table 1. QPD-Original model (*a*) and QPD-simplified model (*b*).

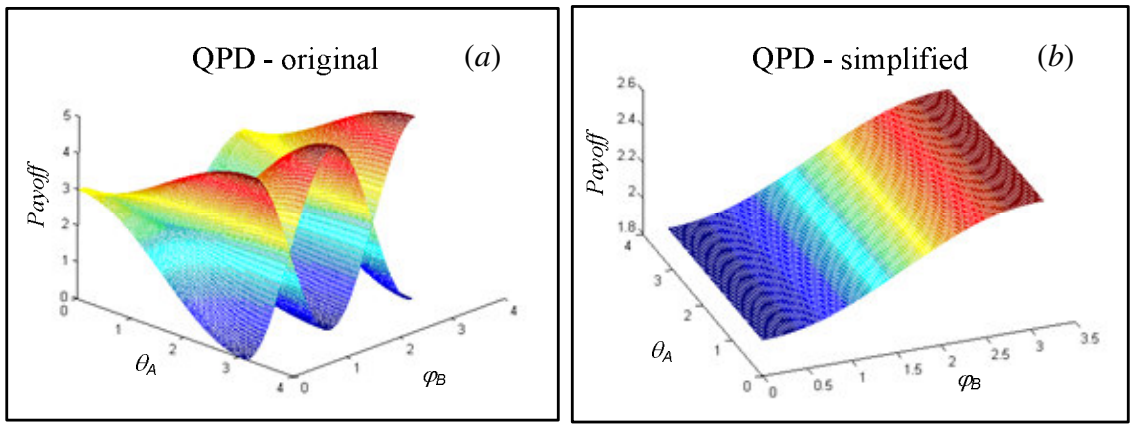

Fig. 4 - Average values of payoffs for players *A* and *B* versus $\theta_A$ and $\varphi_B$, using Table 1. QPD-Original model (*a*) and QPD-simplified model (*b*).

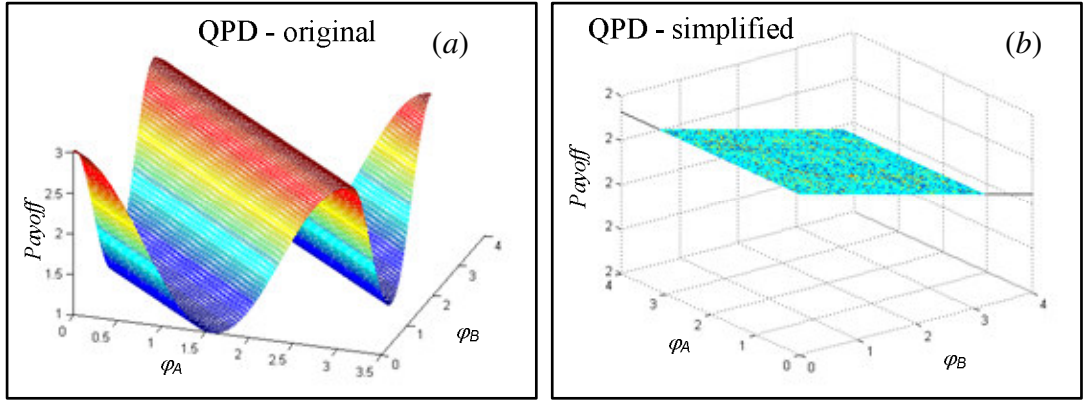

Fig. 5 - Average values of payoffs for players *A* and *B* versus $\varphi_A$ and $\varphi_B$, using Table 1. QPD-Original model (*a*) and QPD-simplified model (*b*).

In Figs. 3-5 one can observe that, in the simplified model, the payoffs' curves for players *A* and *B* are equals and the range of average payoff values is reduced. Obviously, a quantum game with the quantum circuit shown in Fig. 2 is a more restricted game than a quantum game realized with Meyer and Eisert's model. However, as will be shown, it has some interesting practical applications.

## 3. Quantum Multiplayer Games

Multiplayer quantum games were firstly discussed [5, 6] as a generalization of the original Meyer-Eisert's scheme. Here, we consider the multiplayer version of the model for quantum games described in Section 2, in order to create an access controller. We will only consider four players (although the proposed quantum game can be easily extended to a larger number of players), designed by $P_1$, $P_2$, $P_3$ and $P_4$, in a non-cooperative scenario, which means that there is no communication among them and each one desires to win using their own strategies, represented by single qubit gates. The goal of the game is to permit the winner of game to transfer its data to the following stage of the computation. Thus, associated with each player there is a binary string. For the example here considered, these data are shown in Table 2.

| Player | 1 | 2 | 3 | 4 |
|---|---|---|---|---|
| Process' data | $|1001\rangle$ | $|0001\rangle$ | $|1000\rangle$ | $|1111\rangle$ |

Table 2. Players' data for the multiplayer game.

Thus, for player 1, for example, the binary string that it wants to send to the next computation stage is coded in the quantum state $|1001\rangle$.

Now, we consider the entangler gate $J$ that, in opposition to what is commonly considered in the literature (usually $J$ generates a GHZ state), acts generating a four-qubits W state:

$$J|0000\rangle = \frac{(|1000\rangle + |0100\rangle + |0010\rangle + |0001\rangle)}{2}. \tag{8}$$

For the quantum game considered this is convenient because each quantum state of the superposition in (8) represents in a direct way a player. The quantum circuit of the proposed quantum game can be seen in Fig. 6.

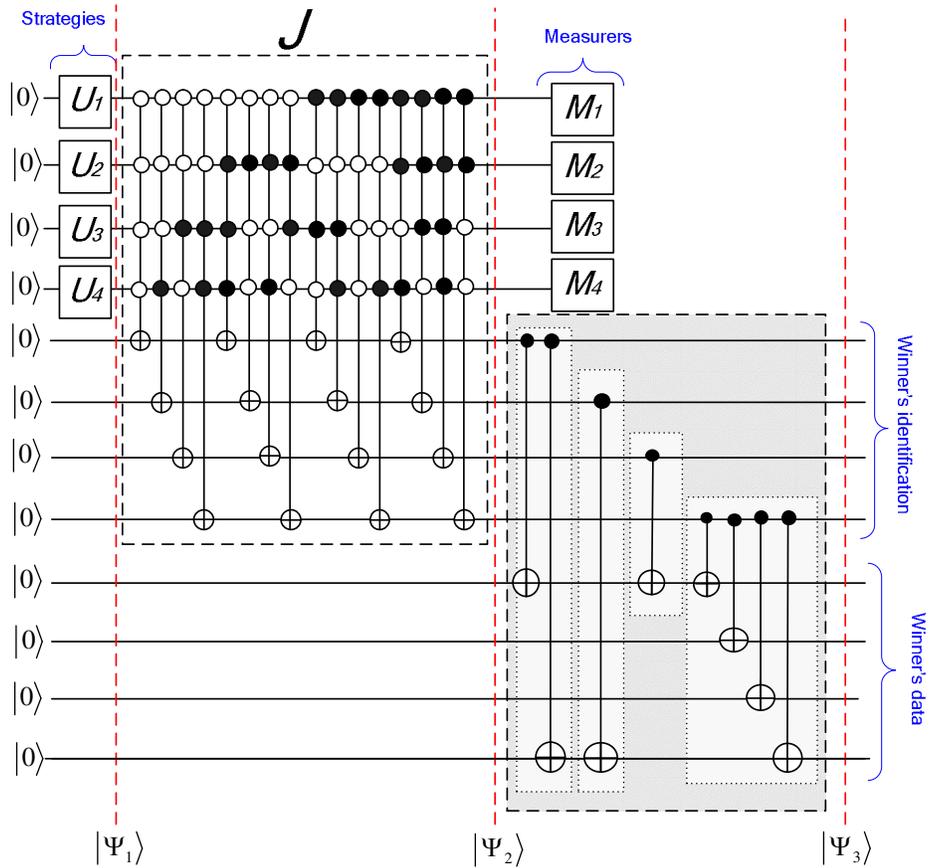

Fig. 6 - Multiplayer game using CNOT grid for access control.

As can be observed in Fig. 6, each player has a grid of CNOTs that are used in accordance to its data. Moreover, one can realize that the game proposed is the multiplayer version of the simplified quantum game model (absence of $J^\dagger$) presented in Section 2. The quantum states in the marked positions in the circuit in Fig. 6 are:

$$|\Psi_1\rangle = U_1|0\rangle \otimes U_2|0\rangle \otimes U_3|0\rangle \otimes U_4|0\rangle \otimes |0000\rangle \otimes |0000\rangle =$$
$$(a_1|0\rangle + b_1|1\rangle) \otimes (a_2|0\rangle + b_2|1\rangle) \otimes (a_3|0\rangle + b_3|1\rangle) \otimes (a_4|0\rangle + b_4|1\rangle) \otimes |0^{\otimes 4}\rangle \otimes |0^{\otimes 4}\rangle \quad (9)$$

$$|\Psi_2\rangle = \begin{bmatrix} (a_1a_2a_3a_4|0000\rangle + a_1b_2b_3b_4|0111\rangle + b_1a_2b_3a_4|1010\rangle + b_1b_2a_3b_4|1101\rangle)|1000\rangle + \\ (a_1a_2a_3b_4|0001\rangle + a_1b_2a_3a_4|0100\rangle + b_1a_2b_3b_4|1011\rangle + b_1b_2b_3a_4|1110\rangle)|0100\rangle + \\ (a_1a_2b_3a_4|0010\rangle + a_1b_2a_3b_4|0101\rangle + b_1a_2a_3a_4|1000\rangle + b_1b_2b_3b_4|1111\rangle)|0010\rangle + \\ (a_1a_2b_3b_4|0011\rangle + a_1b_2b_3a_4|0110\rangle + b_1a_2a_3b_4|1001\rangle + b_1b_2a_3a_4|1100\rangle)|0001\rangle \end{bmatrix} \otimes |0^{\otimes 4}\rangle \quad (10)$$

$$|\Psi_3\rangle = \begin{bmatrix} (a_1a_2a_3a_4|0000\rangle + a_1b_2b_3b_4|0111\rangle + b_1a_2b_3a_4|1010\rangle + b_1b_2a_3b_4|1101\rangle)|1000\rangle|1001\rangle + \\ (a_1a_2a_3b_4|0001\rangle + a_1b_2a_3a_4|0100\rangle + b_1a_2b_3b_4|1011\rangle + b_1b_2b_3a_4|1110\rangle)|0100\rangle|0001\rangle + \\ (a_1a_2b_3a_4|0010\rangle + a_1b_2a_3b_4|0101\rangle + b_1a_2a_3a_4|1000\rangle + b_1b_2b_3b_4|1111\rangle)|0010\rangle|1000\rangle + \\ (a_1a_2b_3b_4|0011\rangle + a_1b_2b_3a_4|0110\rangle + b_1a_2a_3b_4|1001\rangle + b_1b_2a_3a_4|1100\rangle)|0001\rangle|1111\rangle \end{bmatrix}. \quad (11)$$

After the measurement, the winner is known and the quantum state with binary sequence corresponding to it appears at the data bus and the winner identification appears in the identification bus. Observing (10)-(11) we see clearly that the probability of one player to win depends on the strategies chosen by all players.

In a computer, where processes are trying to access the CPU, it is common to establish a set of priorities, since some processes may be more important than others. In order to take this into account, we are going to use a measure to evaluate the priorities of the processes. This measure is simply a variable, named $\varepsilon$, in the interval [0,1]. Thus, a process with total priority has $\varepsilon=1$ (that is, it should always be ran when requested) while a process with none priority (that is, it should never be ran) has $\varepsilon=0$. Thus, we can define a priority scale for the set of processes. Let us assume that player $k$ has priority $\varepsilon_k$. This can be translated to the game saying that player $k$ ($k=1,2,3,4$) has probability $\varepsilon_k$ of being the winner of the game. However, according to (11), these probabilities are:

$$p_1 = |a_1a_2a_3a_4|^2 + |a_1b_2b_3b_4|^2 + |b_1a_2b_3a_4|^2 + |b_1b_2a_3b_4|^2 \quad (12)$$

$$p_2 = |a_1a_2a_3b_4|^2 + |a_1b_2a_3a_4|^2 + |b_1a_2b_3b_4|^2 + |b_1b_2b_3a_4|^2 \quad (13)$$

$$p_3 = |a_1a_2b_3a_4|^2 + |a_1b_2a_3b_4|^2 + |b_1a_2a_3a_4|^2 + |b_1b_2b_3b_4|^2 \quad (14)$$

$$p_4 = |a_1a_2b_3b_4|^2 + |a_1b_2b_3a_4|^2 + |b_1a_2a_3b_4|^2 + |b_1b_2a_3a_4|^2 \quad (15)$$

with the condition $p_1+p_2+p_3+p_4=1$. The Pareto's optimal is obtained when a set of strategies $\{U_1,U_2,U_3,U_4\}$ leads to $p_1=\varepsilon_1$, $p_2=\varepsilon_2$, $p_3=\varepsilon_3$, and $p_4=\varepsilon_4$. For example, let us suppose that $\varepsilon_1=0.4$, $\varepsilon_2=0.3$, $\varepsilon_3=0.2$ and $\varepsilon_4=0.1$.

Using a genetic algorithm with 12-bits chromosomes, population with 100 individuals and 1000 generations, we have found two sets of strategies that lead to Pareto's optimal:

$$\{U_1^1, U_2^1, U_3^1, U_4^1\} = \left\{ \begin{bmatrix} -0.0038 - 0.4920i & 0.8361 + 0.2427i \\ -0.8361 + 0.2427i & -0.0038 + 0.4920i \end{bmatrix}, \begin{bmatrix} -0.2486 - 0.7967i & 0.5318 - 0.1438i \\ -0.5318 - 0.1438i & -0.2486 + 0.7967i \end{bmatrix}, \begin{bmatrix} -0.1978 + 0.0281i & -0.3488 + 0.9157i \\ 0.3488 + 0.9157i & -0.1978 - 0.0281i \end{bmatrix}, \begin{bmatrix} -0.7550 + 0.4130i & 0.1562 - 0.4847i \\ -0.1562 - 0.4847i & -0.7550 - 0.4130i \end{bmatrix} \right\} \quad (16)$$

$$\{U_1^2, U_2^2, U_3^2, U_4^2\} = \left\{ \begin{bmatrix} 0.1769 + 0.3634i & 0.8405 - 0.3607i \\ -0.8405 - 0.3607i & 0.1769 - 0.3634i \end{bmatrix}, \begin{bmatrix} -0.7518 + 0.4023i & 0.3505 + 0.3874i \\ -0.3505 + 0.3874i & -0.7518 - 0.4023i \end{bmatrix}, \begin{bmatrix} -0.2903 + 0.2477i & 0.8993 - 0.2138i \\ -0.8993 - 0.2138i & -0.2903 - 0.2477i \end{bmatrix}, \begin{bmatrix} 0.7454 - 0.3901i & -0.2846 - 0.4597i \\ 0.2846 - 0.4597i & 0.7454 + 0.3901i \end{bmatrix} \right\} \quad (17)$$

Using the set of strategies 1 $\{U_1^1, U_2^1, U_3^1, U_4^1\}$ (player $k$ uses $U_k^1$) we get $p_1$=0.4010, $p_2$=0.2990, $p_3$=0.1935 and $p_4$=0.1065, while using the set of strategies 2 $\{U_1^2, U_2^2, U_3^2, U_4^2\}$ we get $p_1$=0.4009, $p_2$=0.2996, $p_3$=0.1934 and $p_4$=0.1061.

Now, let us suppose the set of priorities $\varepsilon_1$=0.15, $\varepsilon_2$=0.35, $\varepsilon_3$=0.35 e $\varepsilon_4$=0.14. Using again the genetic algorithm with the same features explained in the earlier example, the following Pareto's optimal was found

$$\{U_1^1, U_2^1, U_3^1, U_4^1\} = \left\{ \begin{bmatrix} -0.2432 + 0.8720i & -0.4180 + 0.0762i \\ 0.4180 + 0.0762i & -0.2432 - 0.8720i \end{bmatrix}, \begin{bmatrix} -0.1250 - 0.1227i & -0.9785 + 0.1093i \\ 0.9785 + 0.1093i & -0.1250 + 0.1227i \end{bmatrix}, \begin{bmatrix} 0.3280 - 0.8451i & 0.3140 + 0.2821i \\ -0.3140 + 0.2821i & 0.3280 + 0.8451i \end{bmatrix}, \begin{bmatrix} 0.6939 + 0.1171i & 0.6621 - 0.2578i \\ -0.6621 - 0.2578i & 0.6939 - 0.1171i \end{bmatrix} \right\} \quad (18)$$

Using the strategies given in (18), we obtain $p_1$=0.1548, $p_2$=0.3496, $p_3$=0.3497 and $p_4$=0.1459.

The quantum game circuit must be reconfigurable [7], since data's processes change all the time. Furthermore, one can observe that the strategy set {$H,H,H,H$} is a Nash equilibrium. At last, the main disadvantages of the quantum game in Fig. 6 are the complexity of $J$ and the number of CNOTs used.

## 4. Using the multiplayer quantum game as access controller in a quantum computer

The quantum game proposed is now used to control the access of processes to the CPU in a quantum computer. We are going to illustrate this task with an example. Let us assume that the quantum computer is simply a circuit able to implement a quantum search [8] using a configurable oracle. Let us also assume that our quantum search is used to invert a function, that is, given a function $f$ and a binary string $y$, it has to find $x$ such that $f(x)=y$. This can be achieved using the oracle shown in Fig. 7 [9].

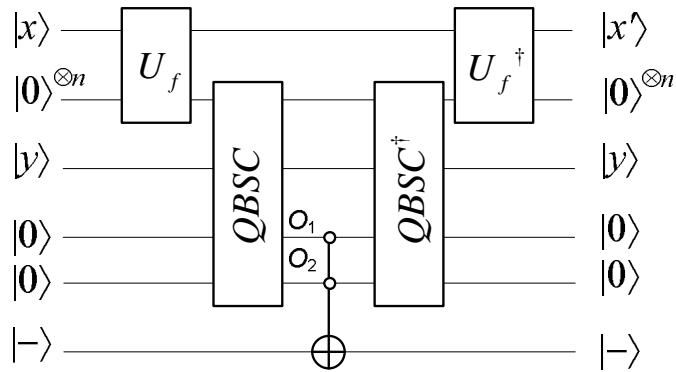

Fig. 7 – Configurable oracle for a quantum search used for inversion of a function $f$.

In Fig. 7, $U_f$ is the unitary operation that implements $f$, $U_f(|x\rangle|0\rangle^{\otimes n})=|x\rangle|f(x)\rangle$, and the circuit QBSC is a quantum bit string comparator and it is shown in Fig. 8 for 4-qubit strings.

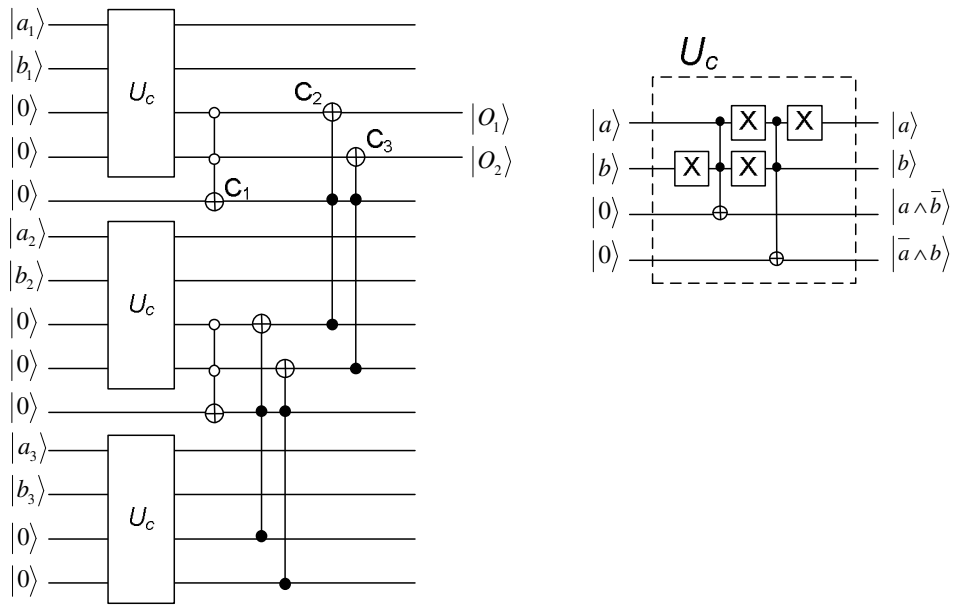

Fig. 8 – Quantum bit string comparator for 4-qubit strings.

In Fig. 7, the strings that will be compared are $|a_1a_2a_3a_4\rangle$ and $|b_1b_2b_3b_4\rangle$. If $a>b$ then $|O_1,O_2\rangle=|10\rangle$, if $a<b$ then $|O_1,O_2\rangle=|01\rangle$ and if $a=b$ then $|O_1,O_2\rangle=|00\rangle$.

Returning to the quantum computer architecture, there are four processes trying to access the quantum CPU in order to run the quantum search. As can be noted in Fig. 7, the state $|x\rangle$ is the database and $|y\rangle$ is a configurable parameter. Putting together the quantum circuits in Figs. 6 and 7, $|y\rangle$ is the quantum state in the data bus of the circuit in Fig. 6. The scheme is shown in Fig. 9.

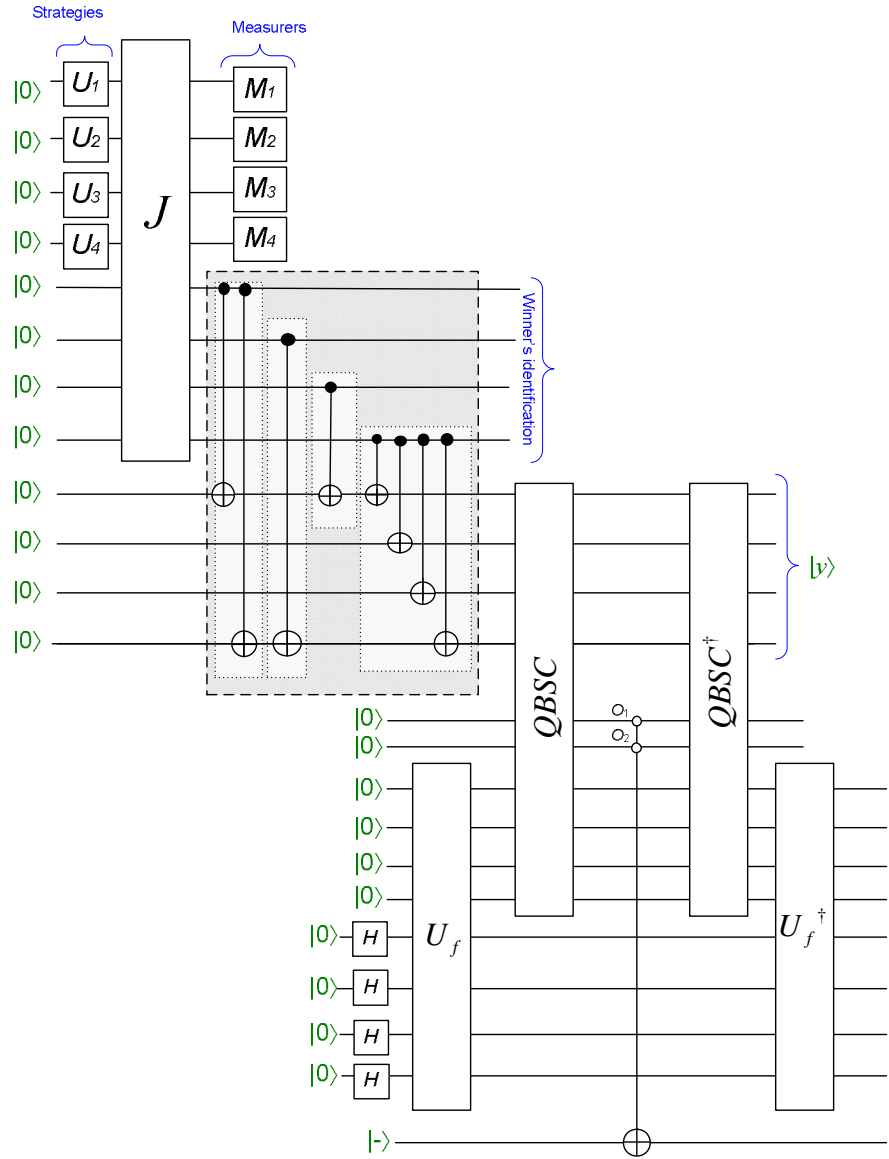

Fig. 9 – Multiplayer quantum game used to control the access to a configurable oracle.

Therefore, when a player wins the game, its associated data is used in order to set the oracle of the quantum search algorithm. Thus, the quantum computer will calculate the inverse of *f* using the winner's data.

## 5. Conclusions

Firstly we have proposed a simplified model for quantum games, showing that, although having a more restrict behavior, it keeps the main characteristic of a quantum game, the interference between strategies. Following, it was presented an access controller based on a multiplayer quantum game using the proposed simplified model. Some Nash and Pareto's equilibriums for the proposed quantum game were numerically found using a genetic algorithm. At last, we showed a simple quantum computer architecture based on the quantum search using the proposed access controller. The quantum game considered had only four players, but a larger version considering a larger number of players is straightforward. Although the proposed quantum game has some complexities as the entangler gate *J* and the number of CNOTs employed, we believe this work is a step towards the construction of an operational system for quantum computers, being quantum games an important tool to be considered.

## Acknowledgements

This work was supported by the Brazilian agency FUNCAP.